\documentclass[prd,reprint,floatfix]{revtex4-1}
\usepackage{graphicx}
\usepackage{float}
\usepackage{wrapfig}
\expandafter\let\csname equation*\endcsname\relax
\expandafter\let\csname endequation*\endcsname\relax
\usepackage{color,soul}
\usepackage{amsmath}
\usepackage{listings}
\usepackage{amssymb}
\usepackage{relsize}
\usepackage{tikz}
\hypersetup{colorlinks=true,citecolor=red,urlcolor=blue}
\usepackage[retainorgcmds]{IEEEtrantools}
\usepackage{pgfplots}
\usetikzlibrary{patterns}
\pgfplotsset{width=10cm, compat=1.10}
\usepgfplotslibrary{fillbetween}
\pgfmathdeclarefunction{surf1}{0}{\pgfmathparse{0.2*sin(x)}}
\pgfmathdeclarefunction{surf2}{0}{\pgfmathparse{-4+0.2*sin(x)}}

\begin{document}

\title[GGN-MIGA]{Characterizing Earth gravity field fluctuations with the MIGA antenna for future Gravitational Wave detectors}

\author{J. Junca$^{1,2}$}
\author{A. Bertoldi$^{1,2}$}
\author{D. O. Sabulsky$^{1,2}$}
\author{G. Lef{\`e}vre$^{1,2}$}
\author{X. Zou$^{1,2}$}
\author{J.-B. Decitre$^{1,3}$}
\author{R. Geiger$^{1,4}$}
\author{A. Landragin$^{1,4}$}
\author{S. Gaffet$^{1,3}$}
\author{P. Bouyer$^{1,2}$}
\author{B. Canuel$^{1,2}$}

\affiliation{$^1$MIGA consortium}
\affiliation{$^2$LP2N, Laboratoire Photonique, Num{\'e}rique et Nanosciences, Universit{\'e} Bordeaux--IOGS--CNRS:UMR 5298, rue F. Mitterrand, F--33400 Talence, France}
\affiliation{$^3$LSBB, Laboratoire Souterrain à Bas Bruit, UNS, UAPV, CNRS:UMS 3538, AMU, La Grande Combe, F--84400 Rustrel, France}
\affiliation{$^4$LNE--SYRTE, Observatoire de Paris, Universit{\'e} PSL, CNRS, Sorbonne Universit{\'e}, 61, avenue de l'Observatoire, F--75014 PARIS, France}

\begin{abstract}
Fluctuations of the earth's gravity field are a major noise source for ground-based experiments investigating general relativity phenomena such as Gravitational Waves (GWs). Mass density variations caused by local seismic or atmospheric perturbations determine spurious differential displacements of the free falling test masses, what is called Gravity Gradient Noise (GGN); it mimics GW effects. This GGN is expected to become dominant in the infrasound domain and must be tackled for the future realization of observatories exploring GWs at low frequency. GGN will be studied with the MIGA experiment, a demonstrator for low frequency GW detection based on atom interferometry - now in construction at the low noise underground laboratory LSBB in France. MIGA will provide precise measurements of local gravity, probed by a network of three free-falling atom test masses separated up to 150~m. We model the effect of GGN for MIGA and use seismic and atmospheric data recorded at LSBB to characterize their impact on the future measurements. We show that the antenna will be able to characterize GGN using dedicated data analysis methods.
\end{abstract}

\maketitle

\section{Introduction}

The first detection of GWs~\cite{Abbott2016} unveiled a new vantage point to study the universe through the observation of phenomena hidden to electromagnetic detectors such as black holes merging~\cite{Abbott2016b}, and later brought to multi-messenger astronomy thanks to the combined electromagnetic and gravitational observation of the collision of a neutron star binary ~\cite{Abbott2017,Abbott2017b}. These novel possibilities define the field of GW astronomy. Whereas the existing GW detectors will keep observing events in their sensitivity window (from a few tens of Hz to 1 kHz~\cite{Martynov2016}), important scientific arguments push to extend the detection bandwidth into lower frequency regimes; a wealth of sources are expected in the infrasound band~\cite{Mandel2018} and tracking signals at low frequency will improve the parameter estimation for events later entering the detection bandwidth of current optical interferometry based GW detectors~\cite{Sesana2016,Graham2018}. The main problem to extend the sensitivity bandwidth of present GW detectors at low frequency is posed by seismic noise~\cite{Martynov2016} and Gravity Gradient Noise~\cite{Harms2015}.

One solution is to develop space based detectors to exploit high quality geodetic motion provided by drag free satellites; that is the case of the laser interferometers eLISA~\cite{Jennrich2009} and DECIGO~\cite{Sato2017}, as well as several proposals relying on ultra-cold atomic sensors~\cite{Norcia2017}, both in the form of Atom Interferometers (AIs) \cite{Yu2011, Graham2013} and atomic clocks~\cite{Kolkowitz2016}. Other approaches relies on improved technologies on earth~\cite{Harms2013} to reduce the impact of low frequency noise sources by improving the mechanical decoupling of the test masses from the environment via enhanced suspensions~\cite{Harms2013} or by opting for freely falling probes which is possible using AIs~\cite{Dimopoulos2008}. Such solutions mitigate the problems created by seismic noise, but do not alter the effect of GGN. 

It has been shown~\cite{Chaibi2016} that the impact of GGN can be strongly reduced by averaging the signals of several distant atom interferometric gradiometers. Such configurations, that require precise GGN models, will be tested with the MIGA experiment~\cite{Canuel2018}, a demonstrator for low frequency GW detection based on atom interferometry and now under construction at the low noise underground laboratory LSBB in France. MIGA will provide precise measurements of the local gravity on a baseline up to 150~m using a network of atom gradiometers. In this article, we model the effect of GGN on MIGA and study possible measurement protocols to validate such models with future data from the antenna.

The article is organized as follows: Sec.~\ref{section1} introduces the sensitivity to GW for a gradiometer based on atom interferometry. We then present projections of the equivalent strain induced by local gravity field fluctuations using seismic (Sec.~\ref{sec:seismic}) and atmospheric (Sec.~\ref{sec:infraS}) data measured on the site where MIGA is being built. In Sec.~\ref{sec:timeint} we then show that MIGA could observe cumulative effects induced by GGN and test the validity of the models used in this article.

\section{GGN and gravity-gradiometry based on AI}\label{section1}
\subsection{Strain limitations from GGN}\label{subsection1-A}
\begin{figure}[ht]
\centering
\resizebox{8.6cm}{!}{
\begin{tikzpicture}%[]
  \draw[thin] (0.5,0) -- (6.5,0);
  \draw[thick] (1.5,-.15) -- (1.5,.15) node at (1.5,-.6) {$X_i$};
  \draw[thick] (5.5,-.15) -- (5.5,.15) node at (5.5,-.6) {$X_j$};
  \shade[top color=red, bottom color=white,opacity=0.5] (0.5,0) rectangle (6.5,-0.15);
  \shade[top color=white, bottom color=red,opacity=0.5] (0.5,0) rectangle (6.5,0.15);  
  \filldraw[fill=white, draw=white] (3.4,-.3) rectangle (4,.3);
  \draw [line width=.35 mm, black ] (3.35,-.2) -- (3.45,.2);
  \draw [line width=.35 mm, black ] (3.95,-.2) -- (4.05,.2);
  \filldraw[fill=black, draw=black] (3.5,0) circle [radius=0.03];
  \filldraw[fill=black, draw=black] (3.7,0) circle [radius=0.03];
  \filldraw[fill=black, draw=black] (3.9,0) circle [radius=0.03];
  \draw[black, very thick] (6.5,-0.5) -- (6.5,0.5);
  \filldraw[fill=cyan, draw=cyan,opacity=0.3] (6.5,-.48) rectangle (6.65,.5);
  \filldraw[fill=blue!80!white, draw=blue!80!white,opacity=0.3] (1.5,0) circle [radius=0.2];
  \filldraw[fill=blue!80!white, draw=blue!80!white,opacity=0.3] (5.5,0) circle [radius=0.2];
  \draw[<->,thick] (1.5,.5) -- (5.5,.5);
  \node at (3.5,1) {$d_{ij}$};
  \node[draw=red,thick,rounded corners,inner sep=2pt] at (0.08,0) {\textcolor{red}{laser}};
  \draw[->] (6.5,0) -- (7,0) node[right] {$X$};
  \draw[thick] (0.5,-.15) -- (0.5,.15) node at (0.5,-.6) {$0$};
  \node at (6.4,-.6) {$L_T$};
\end{tikzpicture}
}
\caption{Gravity gradiometer realized with two AIs placed in $X_{i,j}$, separated by a distance $d_{ij}$, and coherently manipulated by a common laser retro-reflected by a mirror placed at position $L_T$.}
\label{fig:AIgradio}
\end{figure}
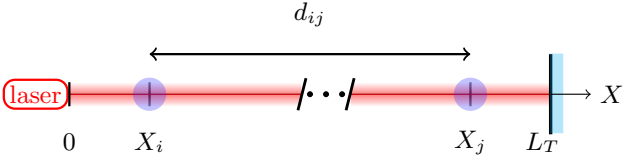

In this section, we consider a single gradiometer using two free-falling atom test masses and calculate the strain sensitivity limitation induced by differential acceleration of the test masses. We consider two correlated AIs placed at positions $X_{i,j}$ along the $x$-axis, created by Bragg-diffraction on the standing wave obtained by the retro-reflection of an interrogating laser (Fig.\ref{fig:AIgradio}).
A three pulse sequence ($\pi /2$--$ \pi$--$ \pi/2$) applied to the two atomic ensembles along the $x$ direction produces the interferometric signal, which is read out as a population unbalance, and depends on the phase difference $\Delta\varphi_{las}$ between the two counter-propagating beams. Considering a large momentum transfer diffraction, i.e. $2n$ photons coherently exchanged during the diffraction process, the output atom phase $\Delta \phi_x(X_i,t)$ of the interferometer at position $X_{i}$ and time $t$ is:
%%%%
\begin{equation}
\Delta \phi_x(X_i,t)=n \int^{\infty}_{-\infty}\Delta\varphi_{las}(X_i,\tau)g'(\tau-t)d\tau+\epsilon(X_i,t) \, ,
\end{equation}
%%%%
where $g'$ is the time derivative of the sensitivity function of the three-pulse AI~\cite{Cheinet08} and $\epsilon(X_i,t)$ is the detection noise related to the projection of the atomic wave-function during the measurement process.
Taking into account the effects of laser frequency noise $\delta\nu(\tau)$, vibration of the retro-reflecting mirror $\delta x_{mir}(\tau)$, GW strain variation $h(\tau)$, and fluctuation of the mean trajectory of the atoms along the laser beam direction induced by the fluctuating local gravity field $\delta x_{at}(X_i,\tau)$, the last equation can be written as~\cite{Chaibi2016}
\begin{align}\label{eq:Deltaphix}
\Delta \phi_x(X_i,t)&=\int^{\infty}_{-\infty}2nk_L\big[\big(\frac{\delta\nu(\tau)}{\nu}+\frac{h(\tau)}{2}\big)(L_T-X_i)\nonumber\\
&+\delta x_{mir}(\tau)-\delta x_{at}(X_i,\tau)\big] g'(\tau-t) d\tau \nonumber\\
&+\epsilon(X_i,t) \, ,
\end{align}
where ${k_L=\frac{2\pi\nu}{c}}$ is the wave number of the interrogation laser. By simultaneously interrogating two AIs with the same laser (Fig.~\ref{fig:AIgradio}), one can cancel, to first order, the motion of the retro-reflecting mirror. The resulting differential phase $\psi(X_i,X_j,t)$ is:
\begin{align}
\psi(X_i,X_j,t)&=\Delta \phi_x(X_i,t)-\Delta \phi_x(X_j,t)\nonumber\\
&=\int^{\infty}_{-\infty}2n k_L\Big[\big(\frac{\delta\nu(\tau)}{\nu}+\frac{h(\tau)}{2}\big)d_{ij} \nonumber\\
&+\delta x_{at}(X_j,\tau)-\delta x_{at}(X_i,\tau)\Big]g'(\tau-t) d\tau\nonumber\\
&+\epsilon(X_i,t)-\epsilon(X_j,t) \, , 
\end{align}
For the sake of clarity we will now neglect the laser frequency noise $\delta\nu(t)$: for MIGA this contribution is expected to be well below the initial sensitivity of the detector~\cite{Canuel2018}. Further, for higher sensitivity configurations, a detector geometry with two orthogonal arms can be used to cancel out the effects of laser frequency noise while preserving sensitivity to $(+)$ polarized GWs. We can write the Power Spectral Density (PSD) of the differential interferometric phase as
\begin{align}\label{eq:Spsi}
\hspace{-0.1cm}
S_{\psi}(d_{ij},\omega)&=(2nk_L)^2 \left[d_{ij}^2\frac{S_h(\omega)}{4}+S_{\Delta x}(d_{ij},\omega)\right]|\omega G(\omega)|^2\nonumber\\
&+2 S_{\epsilon}(\omega) \, ,
\end{align}
given that the detection noise is spatially uncorrelated. 
The term $G(\omega)$ represents the Fourier transform of the sensitivity function of the interferometer to phase variations~\cite{Cheinet08}
\begin{equation}
|\omega G(\omega)|=4 \sin^2\left(\frac{\omega T}{2}\right) \, ,
\end{equation}
where $T$ is the time interval between two successive pulses of the interferometer.

In Eq.~(\ref{eq:Spsi}) the term $S_{\Delta x}(d_{ij},\omega)$ is the PSD of the differential displacement of the atom test masses with respect to the interrogation laser, 
\begin{equation}
S_{\Delta x}(d_{ij},\omega)=\frac{S_{\Delta a_x}(d_{ij},\omega)}{\omega^4} \, ,
\end{equation}
where $S_{\Delta a_x}(d_{ij},\omega)$ is the PSD of the difference of local gravity field between the points $X_i$ and $X_j$ projected along the gradiometer direction, the GGN. This perturbation introduces an atomic phase variation that is indistinguishable from the signal produced by a GW, shown in Eq.~(\ref{eq:Spsi}), and constitutes a limit for the detector that sums up with detection noise.

To quantify and compare the possibility to detect GWs over various noise sources, taking the GW signal as the signal of interest in Eq.~(\ref{eq:Spsi}) and dividing it by the other terms, we obtain the signal to noise ratio (SNR) of the detector. Setting the limit of detectability for an SNR of 1, we define the strain sensitivity of the gradiometer such as the sum
\begin{equation}\label{eq:Sh}
S_h(\omega)= \frac{4S_{\Delta a_x}(d_{ij},\omega)}{\omega^4d^2_{ij}}+\frac{S_{\epsilon}(\omega)}{2(2 n k)^2 d_{ij}^2\sin^4(\frac{\omega T}{2})} \, .
\end{equation}
\subsection{Spatial Correlation of GGN}\label{subsection1-B}

GGN originates from a series of different mechanisms~\cite{Harms2015}, which are associated with different spatial correlation properties. We express GGN as a function of the correlation of the gravity field variations $\delta a_x$ between points $X_i$ and $X_j$ as 
\begin{align}\label{eq:Correl1}
S_{\Delta a_x}(d_{ij},\omega)&=S\{\delta a_x(X_j)-\delta a_x(X_i)\}(\omega)\nonumber\\
&=F\{[\delta a_x(X_j)-\delta a_x(X_i)]\\
&\qquad\quad\ \otimes[\delta a_x^*(X_j)-\delta a_x^*(X_i)]\}(\omega) \nonumber \, ,
\end{align}
where $S\{.\}$, $F\{.\}$ and $\otimes$ denote the PSD, Fourier Transform, and convolution operators respectively. The last equation can be written as
\begin{equation}\label{eq:Ssgnn11}
S_{\Delta a_x}(d_{ij},\omega)= 2C(0,\omega)- C(d_{ij},\omega)-C(-d_{ij},\omega)\, ,
\end{equation}
where $C(d_{ij},\omega)$ is the Fourier Transform of the two point spatial correlation function of the gravity field 
\begin{align}\label{eq:CorellF}
C(d_{ij},\omega)&=F\{\delta a_x(X_j)\otimes\delta a_x^*(X_i)\}(\omega)\nonumber\\
&=F\{\delta a_x(X_j)\}(\omega)\times F\{\delta a_x^*(X_i)\}(\omega)\, .
\end{align}
In the following, we will focus on the two contributions previously identified as the main limiting effects for GW detectors linked to density variations in the surrounding medium resulting from seismic activity~\cite{Saulson,PhysRevD.58.122002} (Sec.~\ref{sec:seismic}) and perturbations of the atmospheric pressure~\cite{Saulson,PhysRevD.97.062003} (Sec.~\ref{sec:infraS}). For both contributions, we will calculate the different gravity correlation functions and project the impact of the associated GGN in terms of strain sensitivity using Eq.~(\ref{eq:Sh}). This study is carried out for the configuration of the MIGA gravity antenna~\cite{Canuel2018}, which will generate gradiometric measurements from a network of three AIs regularly spaced on a baseline of 150~m.

\section{Seismic GGN}\label{sec:seismic}

\begin{figure}[htb]
\includegraphics[width=0.48\textwidth]{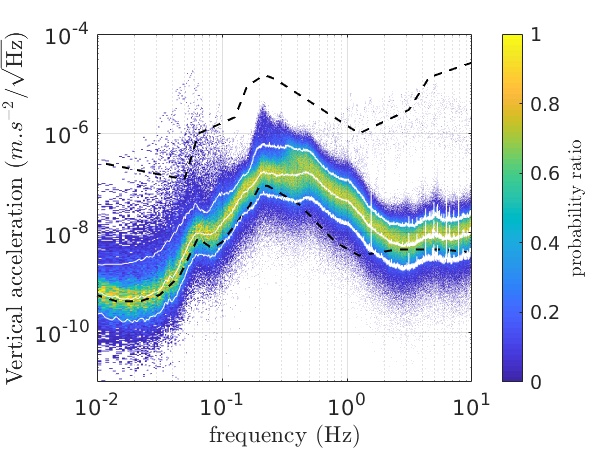}
\includegraphics[width=0.48\textwidth]{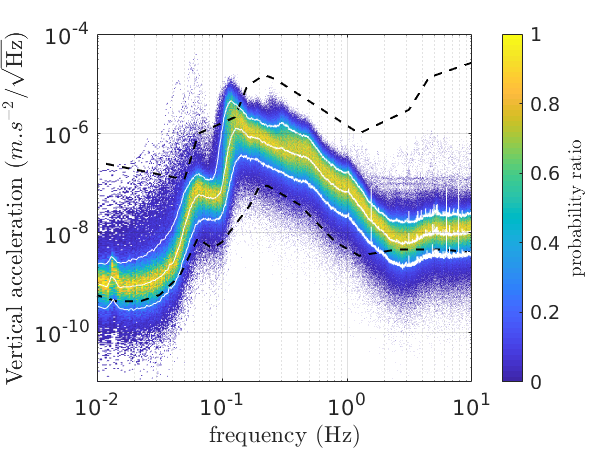}
\caption{Vertical acceleration spectra at LSBB compared to the high and low noise Peterson models~\cite{Peterson1993}, with 10$^{\mathrm{th}}$, 50$^{\mathrm{th}}$ and 90$^{\mathrm{th}}$ percentiles of the histogram represented with white lines. On the top, the spectrum recorded in a quiet month (August 2011); on the bottom, in a noisy month (February 2011).}
\label{fig:seismic}
\end{figure}

The MIGA experiment is now under construction in southern France at the ``Laboratoire Souterrain \`a Bas Bruit" (LSBB), an underground facility located away from major human activities and characterized by very low seismic noise. The seismic conditions within the laboratory are continuously monitored using a network of 6 broadband STS2 seismometers~\cite{Gaffet2009}. Fig.~\ref{fig:seismic} shows typical seismic spectra along the vertical axis, recorded by the station located at the center of the infrastructure (``RAS" station, see~\cite{Gaffet2009}). The data are compared to Peterson's model~\cite{Peterson1993}, which is usually taken as a reference to assess the seismic properties of a given site.

\begin{figure}
    \centering
    \begin{tikzpicture}
    \def\f{-40}
    \def\a{.96}
    \def\b{.28}
  \def\Xo{4}
  \def\Yo{3}
  \draw[->] (-3.5*\a+\Xo,0+\Yo+3.5*\b) -- (3.5*\a+\Xo,0+\Yo-3.5*\b) node[right] {$x$};
  \draw[->] (-2.8+\Xo,-1.5+\Yo) -- (2.8+\Xo,1.5+\Yo) node[right] {$y$};
  \draw[->] (0+\Xo,-2+\Yo) -- (0+\Xo,2+\Yo) node[right] {$z$};

  \draw[->,thick,>=stealth] (0+\Xo,0+\Yo) -- (1.3+\Xo,0+\Yo);
  \node at (1+\Xo,+0.25+\Yo) {$\vec{k}_{R}$};
  \draw [thick,domain=-43:0] plot ({1*cos(\x)+\Xo}, {0.32*sin(\x)+\Yo});
  \node at (1.2+\Xo,-.2+\Yo) {$\phi$};

  \def\Yb{2.5}
  \draw[<->,thick,>=stealth](0+\Xo,0+\Yo)--(0+\Xo,-1+\Yb);
  \node at (-0.25+\Xo,-.6+\Yo) {$h$};
  \draw[->,thick,>=stealth] (0+\Xo,0+\Yo) -- (1+\Xo,-1-\b+\Yb);
  \node at (.85+\Xo,-.8+\Yo) {$\vec{r}_0$};

  \def\ro{-1}
  \draw[thin] (4+\ro*\a,0-1-\ro*\b+\Yb) -- (4+\ro*\a+2.8*\a,0-1-\ro*\b-2.8*\b+\Yb);
  \draw[thick,color=red,opacity=0.5] (4+\ro*\a,0-1-\ro*\b+\Yb+0.03) -- (4+\ro*\a+2.8*\a,0-1-\ro*\b-2.8*\b+\Yb+0.03);
  \draw[thick,color=red,opacity=0.5] (4+\ro*\a,0-1-\ro*\b+\Yb-0.03) -- (4+\ro*\a+2.8*\a,0-1-\ro*\b-2.8*\b+\Yb-0.03);
  \draw[thick] (4+\ro*\a+.3*\a,0-1-\ro*\b-.3*\b+\Yb-0.05) -- (4+\ro*\a+.3*\a,0-1-\ro*\b-.3*\b+\Yb+0.05) node at (4+\ro*\a+.3*\a,0-1-\ro*\b-.3*\b+\Yb-.3) {$X_i$};
  \draw[thick] (4+\ro*\a+2.5*\a,0-1-\ro*\b-2.5*\b+\Yb-0.05) -- (4+\ro*\a+2.5*\a,0-1-\ro*\b-2.5*\b+\Yb+0.05) node at (4+\ro*\a+2.5*\a,0-1-\ro*\b-2.5*\b+\Yb-0.3) {$X_j$};
  \filldraw[fill=blue!80!white, draw=blue!80!white,opacity=0.3] (4+\ro*\a+.3*\a,0-1-\ro*\b-.3*\b+\Yb) circle [radius=0.06];
  \filldraw[fill=blue!80!white, draw=blue!80!white,opacity=0.3] (4+\ro*\a+2.5*\a,0-1-\ro*\b-2.5*\b+\Yb) circle [radius=0.06];
  
  \begin{axis}[zmin=-3,zmax=5,axis lines= none, view={-140}{25}]
    \addplot3[surf, colormap/blackwhite,opacity=.2, samples=20, domain=-450:450] {0.5*cos(x*cos(\f)-y*sin(\f))};
    \addplot3 [name path = xline, draw = none, y domain = -450:450] (y,450,-3);
    \addplot3 [name path = yline, draw = none, y domain = -450:450] (-450,y,-3);
    \addplot3 [name path = xcurve, y domain = -450:450, draw = none]
      (y, 450, {0.5*cos(y*cos(\f)-450*sin(\f))});
    \addplot3 [name path = ycurve, y domain = -450:450, draw = none]
      (-450, y, {0.5*cos(-450*cos(\f)-y*sin(\f))});
    \addplot [color = black!70, draw = none,opacity=.2]
      fill between[of = xcurve and xline];
    \addplot [color = black!30, draw = none,opacity=.2]
      fill between[of = yline and ycurve, reverse = true];
  \end{axis}
\end{tikzpicture}
    \caption{Rayleigh wave modelling: angle and axis definition. $\vec{k}_R$ wave vector of the incident Rayleigh wave, $\phi$ azimuth angle of the wave, $h$ depth of the detector, $\vec{r}_0$ position where the gravity perturbation is calculated, $X_i$ and $X_j$ position of the extremities of the gradiometer along the $X$-axis.}
    \label{fig:notationsRayleigh}
\end{figure}
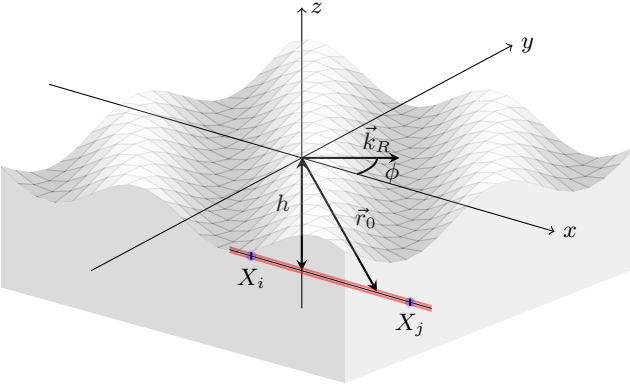

In quiet periods the seismic background activity at LSBB is close to the low Peterson model half of the time. In noisier periods, typically during winter, the seismic background noise is increased with a frequency signature that depends on sea and ocean conditions. The sea and ocean motion accounts for most of the energy that generates the continuous vertical oscillations of the earth~\cite{Ardhuin2015} that propagates as surface seismic waves (Love and Rayleigh waves). Love waves are pure shear waves that don't produce density variations in the medium~\cite{Intro2seismo} and thus don't produce GGN. Therefore, we confine our study to the case of Rayleigh waves and consider the total seismic field as an incoherent sum of monochromatic Rayleigh waves propagating isotropically  over the different angular directions~$\phi$ in the horizontal plane (see Fig.~\ref{fig:notationsRayleigh}).

At an underground depth $h>0$, the gravity potential perturbation $\delta U(\vec{r}_0,t)$ induced by Rayleigh waves propagating in the angular directions~$\phi$ is~\cite{Harms2015}
\begin{align}\label{eq:d_phi_rt}
\delta U(\vec{r}_0,t)&=2\pi G\rho_0\frac{\xi_z(\omega)}{q_z^P-k_R\sqrt{q_z^P/q_z^S}}e^{i(\vec{k}_R\cdot\vec{r}_0-\omega t)}\nonumber\\
&\times\big(-2e^{-hq_z^P}+(1+\sqrt{q_z^P/q_z^S})e^{-hk_R}\big)\, ,
\end{align}
where $\vec{k}_R$ is the wave vector, $\xi_z(\omega)$ is the vertical seismic displacement amplitude, $\rho_0$ is the average medium density and $G$ is the gravitational constant. Note, $k_P$ and $k_S$ are the wave vectors of primary and secondary waves, where we have $q_z^P=\sqrt{k_R^2-k_P^2}$ and $q_z^S=\sqrt{k_R^2-k_S^2}$.

Assuming a linear dispersion with primary wave speed~$\alpha$, secondary wave speed~$\beta$, and Rayleigh wave speed~$c_R$, the gravity perturbation of Eq.~(\ref{eq:d_phi_rt}) determines the longitudinal acceleration on a test mass in~$\vec{r}_0$ as
\begin{align}\label{eq:daxS}
\delta a_x(\vec{r}_0,t)&=-\vec{\nabla}\delta U(\vec{r}_0,t).\vec{e}_x\nonumber\\
&=2\pi G \rho_0 \gamma\xi_z(\omega)e^{i(\vec{k}_R\cdot\vec{r}_0-\omega t)}\nonumber\\
&\times\big[e^{-hk_R}+b(e^{-hq_z^P}-e^{-hk_R})\big]i\cos\phi \, ,
\end{align}
where $\gamma$ and $b$ are dimensionless factors depending on the wave velocity ratios $c_R/\alpha$ and $c_R/\beta$, which themselves depend upon the mechanical properties of the propagating medium:
\begin{equation}
\gamma=\frac{\Big(\frac{1-c_R^2/\alpha^2}{1-c_R^2/\beta^2}\Big)^\frac{1}{4}-1}{\Big(\frac{1-c_R^2/\alpha^2}{1-c_R^2/\beta^2}\Big)^\frac{1}{4}-\sqrt{1-c_R^2/\alpha^2}} \, ,
\end{equation}
\begin{equation}
b=\frac{2}{1-\Big(\frac{1-c_R^2/\alpha^2}{1-c_R^2/\beta^2}\Big)^\frac{1}{4}} \, .
\end{equation}
For a test mass placed at $\vec{r}_0=X\vec{e}_x-h\vec{e}_z$, Eq.~(\ref{eq:daxS}) can be written as:
\begin{equation}
\delta a_x(X,t)=\mathlarger{\kappa}_R(k_R) \xi_z(\omega)e^{-i\omega t}e^{i k_R X\cos\phi}i\cos\phi\, ,
\end{equation}
with
\begin{equation}
\hspace{-0.2cm}\resizebox{.9\hsize}{!}{$\mathlarger{\kappa}_R(k_R):=2\pi G \rho_0 \gamma[e^{-hk_R}+b(e^{-hq_z^P}-e^{-hk_R})]$} \, .
\label{kappaS}
\end{equation}
The spatial correlation of the seismic GGN for an underground gradiometer of baseline  $d_{ij}=X_j-X_i$ can then be obtained from Eq.~(\ref{eq:CorellF}), averaging over all different directions $\phi$,
\begin{align}\label{eq:CRequ}
C^R(d_{ij},\omega)&=\mathlarger{\kappa}_R^2(k_R) S_{\xi_z}(\omega)\left\langle\cos^2\phi e^{ik_Rd_{ij}\cos\phi}\right\rangle_{\phi}\nonumber\\\
&=\frac{1}{2}\mathlarger{\kappa}_R^2(k_R)S_{\xi_z}(\omega) \left[J_0(k_R d_{ij})-J_2(k_R d_{ij})\right]\, ,
\end{align}
with $J_n$ the $n^{\text{th}}$ Bessel function of the first kind. Using Eq.~(\ref{eq:Ssgnn11}) we obtain the associated GGN 
\begin{equation}\label{eq:SsgnnFIN}
S_{\Delta a_x}(d_{ij},\omega)=\mathlarger{\kappa}_R^2(k_R)S_{\xi_z}(\omega) \chi^R_{1}(k_R,d_{ij})\, .
\end{equation}
where $\chi^R_{1}(k_R,d_{ij})$ gathers all correlation effects contributing to the GGN: 
\begin{equation}\label{eq:xsiR}
\chi^R_{1}(k_R,d_{ij})=1-J_0(k_R d_{ij})+J_2(k_R d_{ij})\, ,
\end{equation}
For short distances or low frequencies ($k_R d_{ij}\ll 1$), Eq.~(\ref{eq:xsiR}) is approximately
\begin{equation}
\chi^R_{1}(k_R,d_{ij})\simeq \frac{3}{8}k_R^2d_{ij}^2 .
\label{eq:approxChiR}
\end{equation}
Taken with Eq.~(\ref{eq:Sh}), this means that the strain sensitivity limitations from GGN are independent of the gradiometer length. In the case of MIGA, the maximal length of the gradiometer will be 150~m and the Rayleigh wave speed of the medium surrounding the LSBB is ${c_R\simeq 2.4}$~km/s~\footnote{The speed of the Rayleigh waves $c_R$ was computed~\cite{Intro2seismo} from the speed of the secondary waves $\beta\simeq 2.61$~km/s~\cite{BERES2013} such that $c_R=0.92 \beta$.}, which means that approximation  Eq.~(\ref{eq:approxChiR}) is valid for ${f\ll f_0=\frac{c_R}{2\pi d_{ij}}=2.5}$~Hz. At higher detection frequencies, or for a longer detector, the complete expression in Eq.~(\ref{eq:xsiR}) must be used.

\begin{figure}[htb]
\centering
\includegraphics[width=8.6cm]{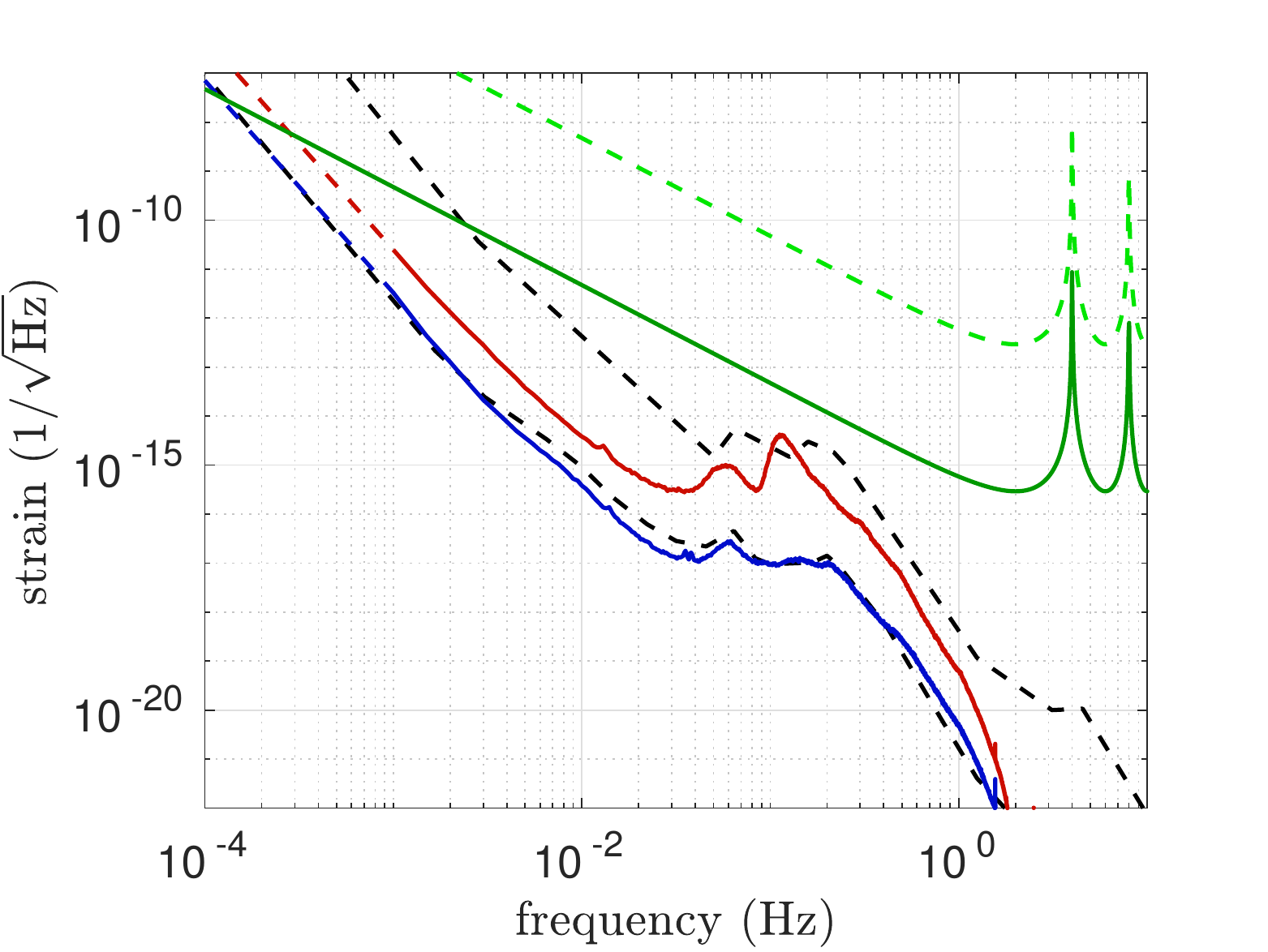}
\caption{In blue (red) seismic GGN projection at LSBB 10$^{\textrm{th}}$ percentile of a quiet month (90$^{\textrm{th}}$ percentile of a noisy month). The low frequency behaviour of the two curves, reported with dashed lines, are extrapolated with a slope of -3.3 in accordance with Peterson's NLNM and NHLM plotted with dashed black lines for comparison. The detection noise for MIGA in its initial and improved configuration, calculated for $2T=0.5$~s, are reported respectively in dashed and plain green lines.}
\label{fig:strainMigaSNN}
\end{figure}

Fig.~\ref{fig:strainMigaSNN} presents the equivalent strain induced by seismic GGN calculated from Eq.~(\ref{eq:SsgnnFIN}). For this projection we used data from the ``RAS" seismic station, considering a depth of $h=300$~m and using the propagation medium parameters that were previously measured at LSBB~\cite{BERES2013}: $\alpha\simeq 4.66$~km/s, $\beta\simeq 2.61$~km/s and $\rho_0=2500$~kg/m$^3$. To take into account the variability of seismic conditions shown in Fig.~\ref{fig:seismic}, the projections are done for the acceleration spectra corresponding to the 90$^{\textrm{th}}$ percentile of a noisy month and the 10$^{\textrm{th}}$ percentile of a quiet month.

We note that the strain limitation from seismic GGN is below the initial, projected shot noise sensitivity of MIGA (light green curve) that comes from the use of standard AI techniques such as 2-photon transitions and a 1 mrad/s detection noise~\cite{Canuel2018}. With an upgraded version of the detector, using large momentum transfer ($2\times 100$ photon transitions) and an enhanced detection noise of $0.1$~mrad/s, we see in Fig.~\ref{fig:strainMigaSNN} that seismic GGN hardly starts to be detectable in the mHz range. Nevertheless, tests of GGN models with this improved detector, referred to in the following as iMIGA, will be possible considering long term integration of the antenna data that will be described in Sec.~\ref{sec:timeint}.

\section{Infrasound GGN}\label{sec:infraS}
We now consider the gravity gradient noise induced by adiabatic pressure variations propagating as sound waves in the atmosphere. We thus consider that relative density and pressure variations $\delta \rho/\rho_0$ and $\delta p/{p_0}$ within the atmosphere are linked by the adiabatic index $\gamma$
\begin{equation}\label{eq:adiab}
\gamma \frac{\delta \rho(\vec{r},t)}{\rho_0}=\frac{\delta p(\vec{r},t)}{p_0} \, .
\end{equation}
The gravity potential perturbation $\delta^3 U(\vec{r}_0,\vec{r},t)$ at $\vec{r}_0$ induced by a single acoustic plane wave of wave vector $\vec{k}$ propagating in an infinitesimal volume $d^3\vec{r}$ centred in $\vec{r}$ is then:
\begin{equation}\label{eq:gravpotinfra}
\delta^3 U(\vec{r}_0,\vec{r},t)=-\frac{G \rho_0}{\gamma p_0}\delta p(\omega)\frac{e^{i(\vec{k}.\vec{r}-\omega t)}}{|\vec{r}-\vec{r}_0|}d^3\vec{r} \, .
\end{equation}
It must be noted that atmospheric density perturbations are not limited to the phenomena described by Eq.~(\ref{eq:adiab}); we limit our study to ``stationary'' perturbations discarding transient perturbations induced by shock waves or airborne objects. Furthermore, given the average depth of the MIGA detector (about 300~m), gravity perturbations from atmospheric temperature fluctuations are believed to be much smaller~\cite{Creighton2008,Harms2015}. Nevertheless a new formalism valid for the frequency window of interest here (10~mHz--1~Hz) would be needed to rule out this contribution definitively.

\begin{figure}
    \centering
    \begin{tikzpicture}
    \def\f{-40}
    \def\a{.96}
    \def\b{.28}
  \begin{axis}[zmin=-3,zmax=5,axis lines= none, view={-140}{25}]
      \addplot3[surf, colormap/blackwhite,opacity=.1, samples=10, domain=-450:450] {0*cos(x*cos(\f)-y*sin(\f))};
    \addplot3 [name path = xline, draw = none, y domain = -450:450] (y,450,-3);
    \addplot3 [name path = yline, draw = none, y domain = -450:450] (-450,y,-3);
    \addplot3 [name path = xcurve, y domain = -450:450, draw = none]
      (y, 450, {0});
    \addplot3 [name path = ycurve, y domain = -450:450, draw = none]
      (-450, y, {0});
    \addplot [color = black!70, draw = none,opacity=.2]
      fill between[of = xcurve and xline];
    \addplot [color = black!30, draw = none,opacity=.2]
      fill between[of = yline and ycurve, reverse = true];
  \end{axis}
  \def\Xo{4}
  \def\Yo{3}
  \draw[->] (-3.5*\a+\Xo,0+\Yo+3.5*\b) -- (3.5*\a+\Xo,0+\Yo-3.5*\b) node[right] {$x$};
  \draw[->] (-2.8+\Xo,-1.5+\Yo) -- (2.8+\Xo,1.5+\Yo) node[right] {$y$};
  \draw[->] (0+\Xo,-2+\Yo) -- (0+\Xo,3+\Yo) node[right] {$z$};
  \draw [thick,domain=0:180] plot ({2.5*cos(\x)+\Xo}, {2.5*sin(\x)+\Yo});
  \draw [dashed,domain=0:180] plot ({2.5*cos(\x)+\Xo}, {0.8*sin(\x)+\Yo});
  \draw [thick,domain=-180:0] plot ({2.5*cos(\x)+\Xo}, {0.8*sin(\x)+\Yo});
  \node at (.8+\Xo,+0.25+\Yo) {$\vec{k}_{\varrho}$};
  \draw [thick,domain=-43:0] plot ({1*cos(\x)+\Xo}, {0.32*sin(\x)+\Yo});
  \node at (1.15+\Xo,-.15+\Yo) {$\phi$};
  \draw[->,very thick,>=stealth] (-1+\Xo,1.8+\Yo) -- (0+\Xo,0+\Yo);
  \draw[->,very thick,>=stealth] (0+\Xo,0+\Yo) -- (1.15+\Xo,1.5+\Yo);
  \draw[->,thick,>=stealth] (0+\Xo,0+\Yo) -- (1.15+\Xo,0+\Yo);
  \draw[dashed] (1.15+\Xo,1.5+\Yo) -- (1.15+\Xo,\Yo);
  \node at (-1+\Xo,1.1+\Yo) {$\vec{k}_I$};
  \draw [thick,domain=90:120] plot ({1.2*cos(\x)+\Xo}, {1.2*sin(\x)+\Yo});
  \node at (-.4+\Xo,1.4+\Yo) {$\theta$};
  \draw [thick,domain=-180:-100] plot ({\x/78+.29+\Xo}, {-\x/50+0.3*sin(8*\x-40)+\Yo});
  \draw [thick,domain=90:180] plot ({\x/78+\Xo}, {\x/60+0.3*sin(8*\x)+\Yo});
  \def\Yb{2.6}
  \draw[<->,thick,>=stealth](0+\Xo,0+\Yo)--(0+\Xo,-1+\Yb);
  \node at (-0.25+\Xo,-.6+\Yo) {$h$};
  \draw[->,thick,>=stealth] (0+\Xo,0+\Yo) -- (1+\Xo,-1-\b+\Yb);
  \node at (.9+\Xo,-.95+\Yo) {$\vec{r}_0$};
  \draw[->,thick,>=stealth] (0+\Xo,-1+\Yb) -- (1*\a+\Xo,-1-\b+\Yb);
  \node at (.5+\Xo,-1.45+\Yb) {$\vec{\varrho}_0$};
  \def\ro{-1}
  \draw[thin] (4+\ro*\a,0-1-\ro*\b+\Yb) -- (4+\ro*\a+2.8*\a,0-1-\ro*\b-2.8*\b+\Yb);
  \draw[thick,color=red,opacity=0.5] (4+\ro*\a,0-1-\ro*\b+\Yb+0.03) -- (4+\ro*\a+2.8*\a,0-1-\ro*\b-2.8*\b+\Yb+0.03);
  \draw[thick,color=red,opacity=0.5] (4+\ro*\a,0-1-\ro*\b+\Yb-0.03) -- (4+\ro*\a+2.8*\a,0-1-\ro*\b-2.8*\b+\Yb-0.03);
  \draw[thick] (4+\ro*\a+.3*\a,0-1-\ro*\b-.3*\b+\Yb-0.05) -- (4+\ro*\a+.3*\a,0-1-\ro*\b-.3*\b+\Yb+0.05) node at (4+\ro*\a+.3*\a,0-1-\ro*\b-.3*\b+\Yb-.3) {$X_i$};
  \draw[thick] (4+\ro*\a+2.5*\a,0-1-\ro*\b-2.5*\b+\Yb-0.05) -- (4+\ro*\a+2.5*\a,0-1-\ro*\b-2.5*\b+\Yb+0.05) node at (4+\ro*\a+2.5*\a,0-1-\ro*\b-2.5*\b+\Yb-0.3) {$X_j$};
  \filldraw[fill=blue!80!white, draw=blue!80!white,opacity=0.3] (4+\ro*\a+.3*\a,0-1-\ro*\b-.3*\b+\Yb) circle [radius=0.06];
  \filldraw[fill=blue!80!white, draw=blue!80!white,opacity=0.3] (4+\ro*\a+2.5*\a,0-1-\ro*\b-2.5*\b+\Yb) circle [radius=0.06];
\end{tikzpicture}
    \caption{Sound wave modelling: angle and axis definition. $\vec{k}_I$ is the wave vector of the incident plane wave,  $\vec{k}_{\varrho}$ being its projection onto the horizontal plane, $\theta$ and $\phi$ the incidence and azimuth angle of the wave, $h$ the depth of the detector, $\vec{r}_0$ the position where the gravity perturbation is calculated, $\vec{\varrho}_0$ being its projection onto the horizontal plane, and $X_i$, $X_j$ the positions of the extremities of the gradiometer along the $X$-axis.}
    \label{fig:notations}
\end{figure}
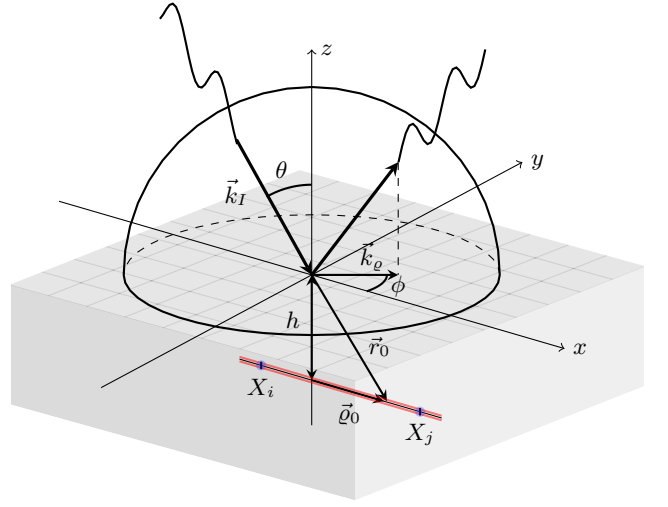

We now consider that acoustic perturbations inside the atmosphere are described by an isotropic superposition of acoustic plane waves. And that  each wave propagating along the direction ($\theta$,$\phi$) with wave-vector $\vec{k}_I$ and velocity $v_s=343$~m/s is totally reflected on the earth's surface (see Fig.~\ref{fig:notations}).
 To obtain the total gravity perturbation $\delta U(\vec{r}_0,t)$ induced on a point situated underground by such a wave, the contributions of the incident and reflected waves given by Eq.~(\ref{eq:gravpotinfra}) must be integrated over the whole hemisphere. The result can be found in ref~\cite{Harms2015},
\begin{equation}
\delta U(\vec{r}_0,t)=-4\pi\frac{G\rho_0}{\gamma p_0}e^{i(\vec{k}_{\varrho}\cdot\vec{\varrho}_0-\omega t)}\frac{\delta p(\omega)}{k_{I}^2}e^{-h k_I\sin\theta} \, .
\label{eq:deltaPhiI}
\end{equation}
Where $\vec{\varrho}_0$ is the projection of $\vec{r}_0$ onto the horizontal plane. This gravity perturbation yields the acceleration:
\begin{multline}
\delta \vec{a}(\vec{r}_0,t)=-\frac{4\pi G\rho_0}{\gamma p_0}e^{i(\vec{k}_{\varrho}\cdot\vec{\varrho}_0-\omega t)}\\
\times\frac{\delta p(\omega)}{k_{I}}e^{-h k_I\sin\theta}\sin \theta
\begin{pmatrix}
 i \cos\phi \\
 i \sin\phi \\
 -1
\end{pmatrix} \, .
\end{multline}
The acceleration created along the gradiometer direction at position $\vec{r}_0=X\vec{e}_x-h\vec{e}_z$ is given by Eq.~(\ref{eq:delta_axI}): 
\begin{multline}\label{eq:delta_axI}
\delta a_x(X,t)=\mathlarger{\kappa}_I(k_I) \delta p(\omega) e^{-i\omega t}e^{i k_I X\sin\theta\cos\phi}\\
\times e^{-h k_I\sin\theta}\sin \theta i\cos\phi  \, ,
\end{multline}
where $\mathlarger{\kappa}_I(k_{I})=\frac{4\pi G\rho_0}{\gamma p_0 k_{I}}$. The spatial correlation function of the atmospheric GGN over the gradiometer can then be obtained from Eq.~(\ref{eq:CorellF}) by averaging over all the different ($\theta$,$\phi$) directions ($\theta \in [0 ; \frac{\pi}{2}]$ and $\phi \in [0 ; 2\pi]$)
\begin{multline}\label{eq:CxxI}
C^I(d_{ij})=\mathlarger{\kappa}_I^2(k_I) S_{\delta p}(\omega) \\
\times\left\langle e^{ik_I d_{ij}\sin\theta\cos\phi}e^{-2hk_I\sin\theta}\sin^2\theta\cos^2\phi \right\rangle_{\theta,\phi} \, .
\end{multline}
From Eq.~(\ref{eq:Ssgnn11}), the infrasound GGN is then
\begin{align}\label{eq:SIGGN2}
S_{\Delta a_x}(d_{ij},\omega)&=\mathlarger{\kappa}_I^2(k_{I})S_{\delta p}(\omega) \chi^I_1(k_{I},d_{ij})  \, ,
\end{align}
where $\chi^I_1(k_I,d_{ij})$ gathers all correlation effects contributing to the GGN 
\begin{multline}\label{eq:chiI}
\chi^I_1(k_I,d_{ij})=\big\langle 2(1-\cos(k_I d_{ij}\sin\theta\cos\phi))\\
 \times e^{-2hk_I\sin\theta}\sin^2\theta\cos^2\phi \big\rangle_{\theta,\phi} \, .
\end{multline}
For $k_Id_{ij}\ll 1$ and $2hk_I\ll1$, a low frequency approximation can be obtained:
\begin{equation}
\chi^I_1(k_I,d_{i j})\simeq \Big(\frac{3}{8}\Big)^2k_I^2 d_{i j}^2\, .
\label{eq:approxChiI}
\end{equation}
For the parameters of MIGA, this approximation stands for $f=v_s/2\pi d_{ij}\ll 0.1$~Hz. In contradiction with the seismic case, this means that around its maximum sensitivity (${f\simeq 2}$~Hz), the full expression Eq.~(\ref{eq:chiI}) has to be used and the strain limitation induced by atmospheric GGN will depend on the distance between the test masses.
\begin{figure}[htb]
\centering
\includegraphics[width=8.6cm]{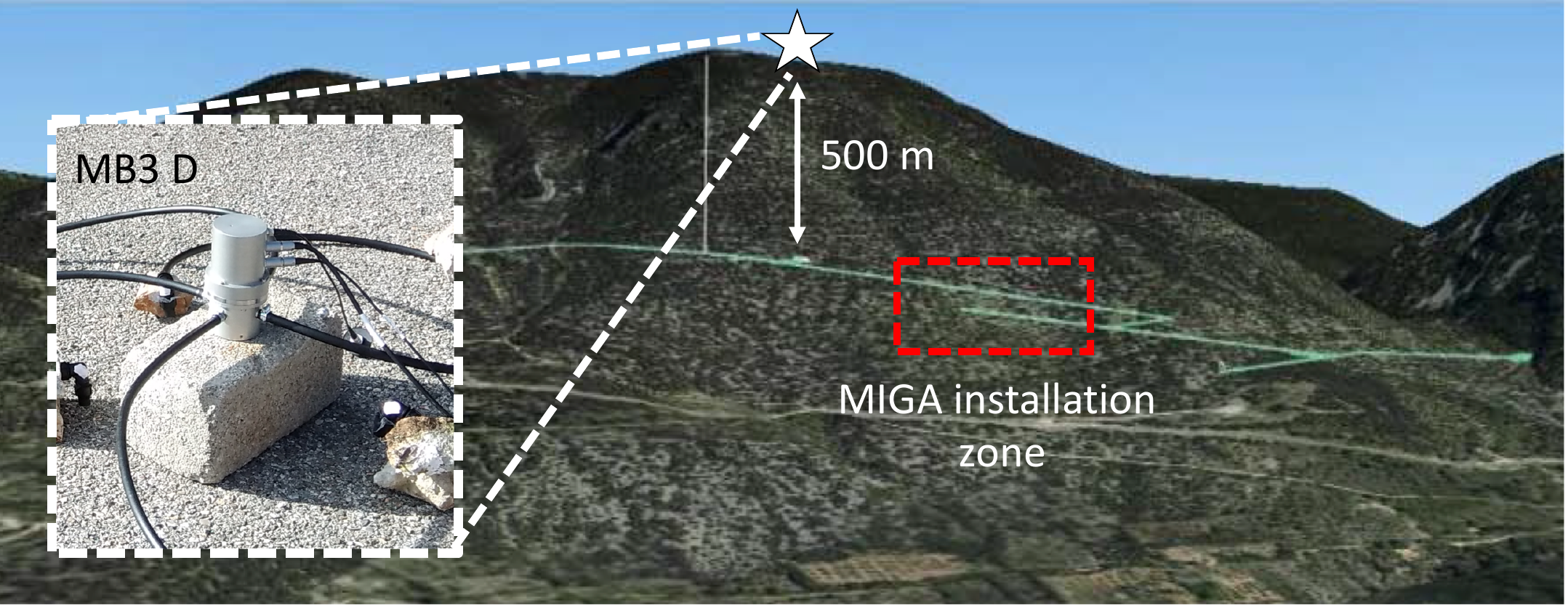}
\caption{Configuration of the pressure measurements at LSBB: a MB3~D microbarometer (white star) was installed on the surface about 500~m above the galleries of the LSBB  (green lines) and close to the future installation zone of MIGA (red box).}
\label{fig:mb3a}
\end{figure}

\begin{figure}[htb]
\includegraphics[width=0.48\textwidth]{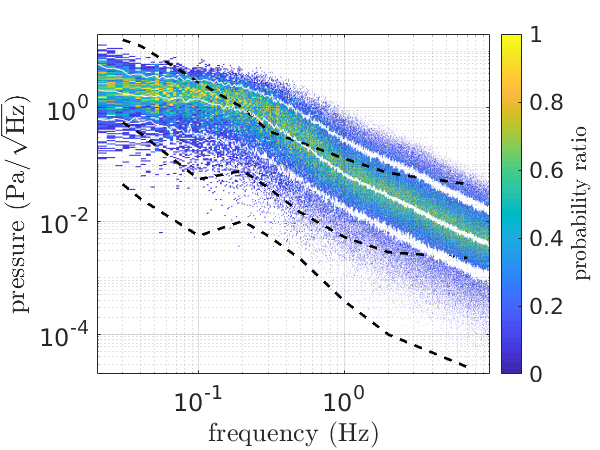}
\includegraphics[width=0.48\textwidth]{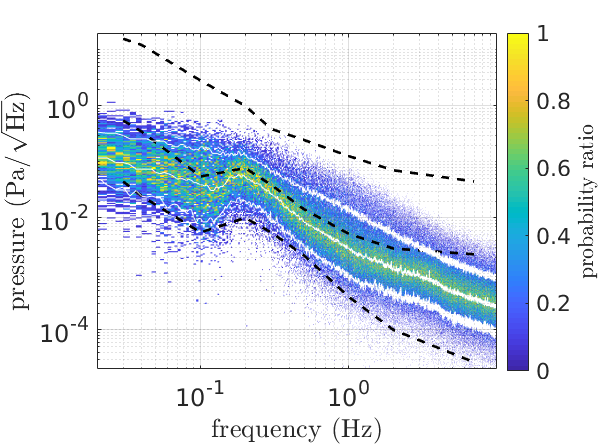}
\caption{Histogram of the outside pressure variations 500~m above the future MIGA galleries for two weather conditions (5~m/s of average wind velocity for the top plot, no wind for the bottom plot) with 10$^{\textrm{th}}$, 50$^{\textrm{th}}$ and 90$^{\textrm{th}}$ percentile in white and Bowman's low, mid and high models in dashed black.}
\label{fig:vestale}
\end{figure}

\begin{figure}[htb]
\centering
\includegraphics[width=8.6cm]{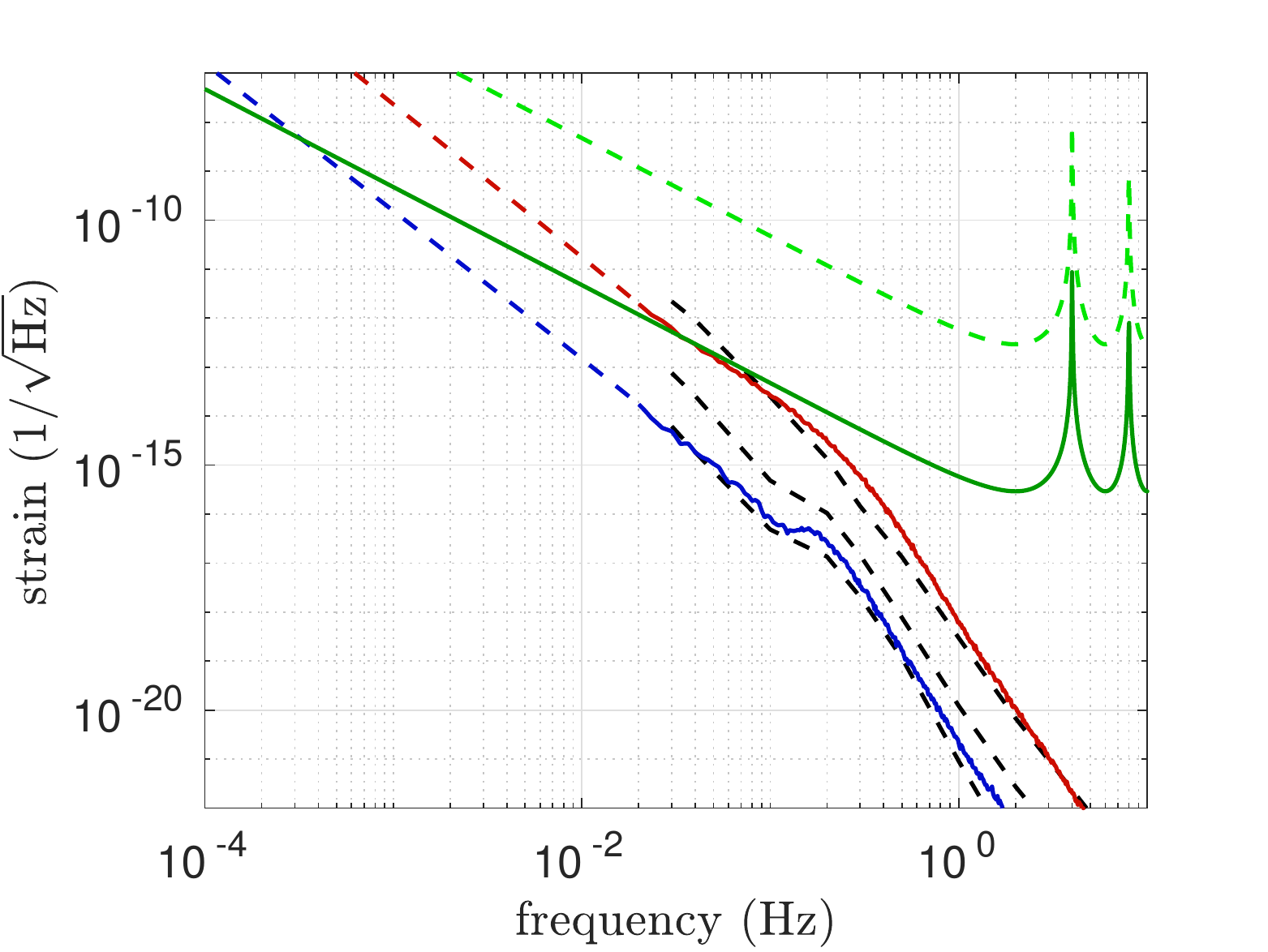}
\caption{MIGA strain sensitivity to infrasound GGN, with data taken from the Bowman model~\cite{Bowman2005} (dashed black) and measured on-site at LSBB (10$^{\textrm{th}}$ percentile of a calm period in blue, and 90$^{\textrm{th}}$ percentile of a windy period in red) with a low frequency extrapolation such that ${S(\delta p;\omega)=S_0 \omega^{-2.2}}$ in accordance with low frequency results reported in~\cite{Marty2010} (dashed blue and red lines). The detection noises for MIGA in its initial and improved configurations, calculated for $2T=0.5$~s, are reported respectively in dashed and plain green lines.}
\label{fig:strainMigaINN}
\end{figure}

Using Eq.~(\ref{eq:SIGGN2}), we project the strain limitation induced by atmospheric GGN on future measurements by MIGA. To determine the local pressure spectrum, we conducted an on-site atmospheric pressure measurement campaign. We used an MB3~D microbarometer from Seismo Wave~\cite{seismowave}, able to resolve 1.75 mPa RMS in the 0.02--4~Hz window, with a self noise at least 10~dB below Bowman's Low Noise Model~\cite{Bowman2005}. The sensor was installed at about 500~m above the galleries of the LSBB and close to the future installation zone of MIGA inside the laboratory (see Fig~\ref{fig:mb3a}).
The measurements were conducted during winter, covering a series of different weather conditions. To filter out wind generated pressure noise~\cite{Alcoverro2005} that can increase pressure variations by several orders of magnitude, we used a simple wind noise attenuator composed of a set of four tubes (1.5~m in length) pointing in the four directions. Fig.~\ref{fig:vestale} shows the amplitude spectral density of pressure variations as histograms for two 12-hour periods: a calm period with no wind and a noisier period associated with a light wind of about 5~m/s. 
A projection of the equivalent strain induced by infrasound GGN corresponding to these measurements is presented in Fig.~\ref{fig:strainMigaINN}. 
Under 0.02~Hz, the projection is performed using an extrapolation of the measurements using a pressure noise model derived from the International Monitoring System stations~\cite{Marty2010}. We note that with the iMIGA configuration the infrasound GGN in noisy weather conditions starts to be detectable below 100~mHz. Improvements upon MIGA could be used to develop more accurate infrasound GGN models that will be required for the future realization of low frequency GW detectors. In this regard, we would like to draw the attention of the reader on the limitations of the current model presented here: the projection uses a pressure noise recorded with a single station which potentially includes a strong contribution that is non-coherent on the scale of a large detector, and thus cannot be modelled by plane waves. Such contributions would need to be estimated using multiple measurement stations in an antenna configuration like that of Matoza \textit{et al.}~\cite{Matoza2013}, for example; It could then be accurately accounted for in the projection model.

\section{Testing GGN models with MIGA}\label{sec:timeint}
\begin{figure}[htb]
\centering
\includegraphics[width=8.6cm]{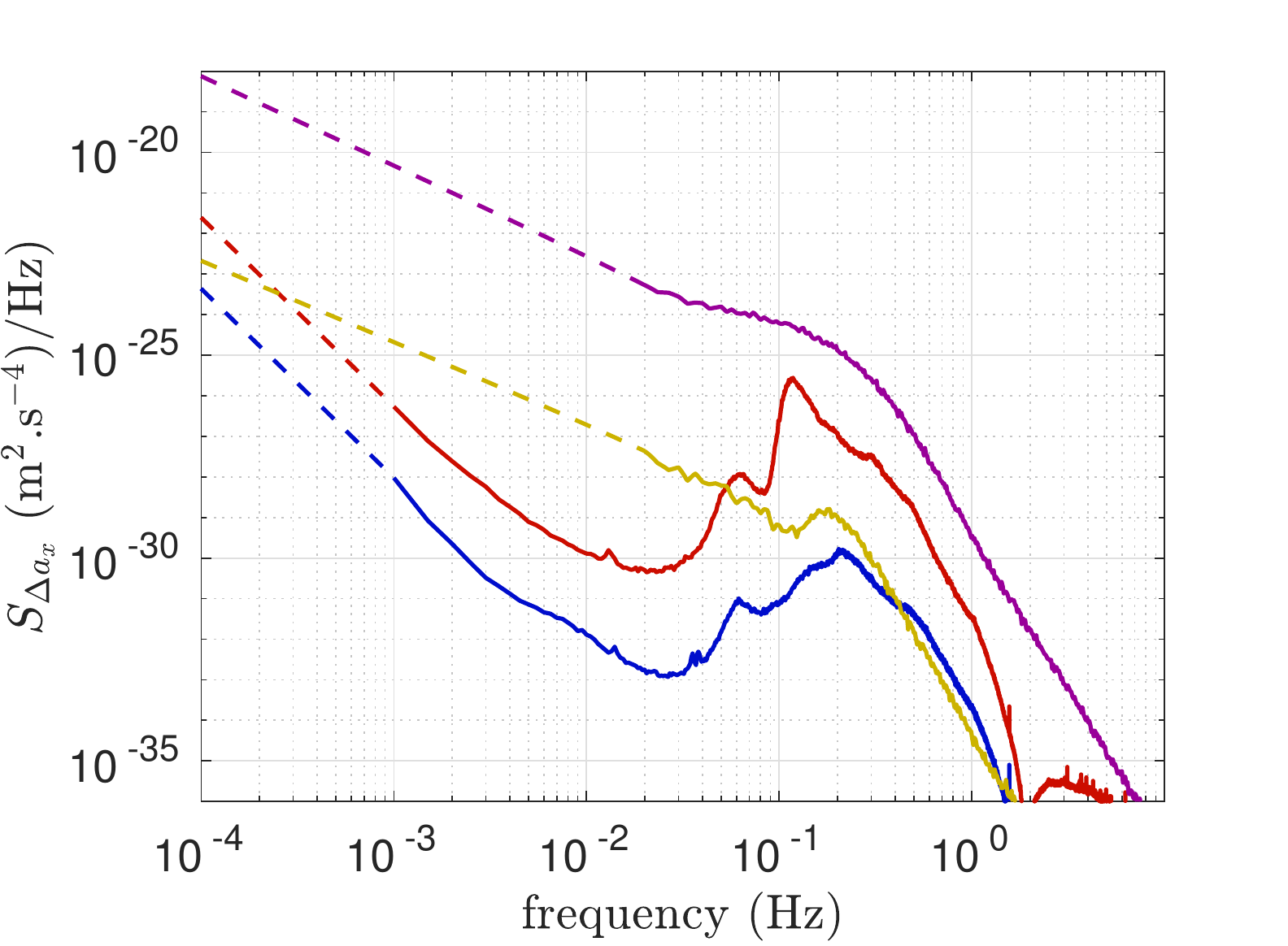}
\caption{Projection of the differential acceleration induced by GGN, calculated using infrasound and seismic data and their low frequency extrapolations. Seismic 10$^{\textrm{th}}$ percentile of a quiet month (90$^{\textrm{th}}$ of a noisy month) in blue (red), and infrasound 10$^{\textrm{th}}$ percentile of a calm period (90$^{\textrm{th}}$ percentile of a windy period) in yellow (purple).}
\label{fig:GGN}
\end{figure}

In the previous sections we have shown that enhanced versions of MIGA could enable GGN spectral analysis. However, due to SNR limitations, this method will be mainly limited to studies of atmospheric GGN, for frequencies under 100~mHz. A large fraction of the projected GGN power spectral density remains under the detection limit set by atom shot noise. Such background noise below the sensitivity threshold can contribute to measurable variations of the atomic phase after integration over an extended period of time. In this section, we present this data analysis method and calculate the Allan variance of the averaged atomic phase from GGN. With the notations introduced in Sec.~\ref{section1}, the gradiometer configuration gives a differential interferometric phase
\begin{align}\label{eq:gradiodiff}
\psi(X_i,X_j,t)=&\Delta \phi_x(X_i,t)\mathsmaller{-}\Delta \phi_x(X_j,t)\nonumber\\
=&n\int_{-\infty}^{+\infty}g'(\tau-t)(\Delta\varphi_{las}(X_i,\tau) \nonumber\\
&-\Delta\varphi_{las}(X_j,\tau))d\tau\, .
\end{align}
Considering that the gradiometer is operated sequentially with a cycling time $T_c$ and that we average $m$ consecutive measurements, we can obtain from ref.~\cite{Cheinet08} the Allan variance of the differential interferometric phase averaged on a time $mT_c$
\begin{equation}
\resizebox{.86\hsize}{!}{$\displaystyle\sigma_{\psi}^2(mT_c)=\frac{n^2}{2m^2}\int_0^{+\infty}\frac{4\sin^4(m\omega T_c/2)}{\sin^2(\omega T_c/2)}|\omega G(\omega)|^2S_{\Gamma}(\omega)\frac{d\omega}{2\pi}
$} \, ,
\end{equation}
where $S_{\Gamma}(\omega)$ is the PSD of the differential laser phase noise $\Delta\varphi_{las}(X_i,\tau)-\Delta\varphi_{las}(X_j,\tau)$. Considering that this noise is coming from acceleration of the atoms induced by GGN yields $S_{\Gamma}(\omega)=\left(\frac{2 k_L}{\omega^2}\right)^2S_{\Delta a_x}(\omega)$, the Allan variance can be expressed as
\begin{equation}\label{eq:allanVar}
\sigma_{\psi}^2(mT_c)=\int_0^{+\infty}H_m(\omega)S_{\Delta a_x}(\omega)\frac{d\omega}{2\pi}\, ,
\end{equation}
where $H_m$ is the transfer function from GGN to averaged differential atomic phase: 
\begin{equation}\label{eq:transferGGN}
H_m(\omega)=\frac{8k_L^2n^2}{m^2}\frac{\sin^4(m\omega T_c/2)}{\sin^2(\omega T_c/2)}\frac{|G(\omega)|^2}{\omega^2}\, .
\end{equation}
Fig.~\ref{fig:GGN} shows the GGN power spectral density $S_{\Delta a_x}$ calculated for the largest gradiometer ($d_{ij}=150$ m) available in MIGA. 
This PSD is obtained from the data and projection models presented in Sec.~\ref{sec:seismic} and~\ref{sec:infraS}. We observe that under 100~mHz, the GGN grows rapidly as $f^{-2.2}$ and $f^{-4.6}$ for atmospheric and seismic contributions respectively.

\begin{figure}[htb]
\includegraphics[width=8.6cm]{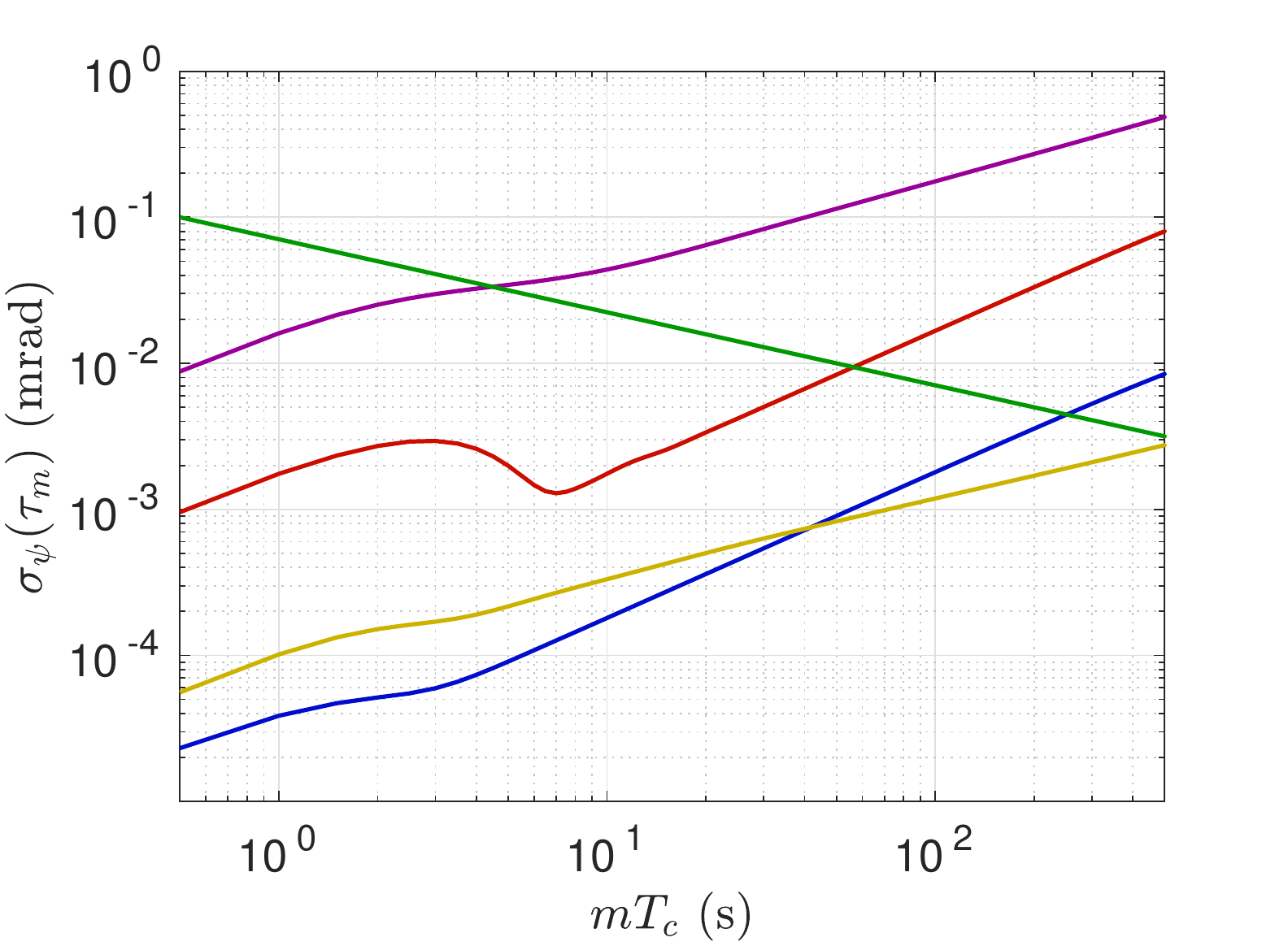}
\caption{Allan standard deviation $\sigma_{\psi}(\tau_m)$ versus the measuring time $m \, T_c$. The cycle time is taken equal to the interrogation time: ${2 T=T_c=0.5}$~s. The calculation adopts the GGN acceleration spectra in Fig.~\ref{fig:GGN}: seismic 10$^{\textrm{th}}$ percentile of a quiet month (90$^{\textrm{th}}$ of a noisy month) in blue (red), and infrasound 10$^{\textrm{th}}$ percentile of a calm period (90$^{\textrm{th}}$ percentile of a windy period) in yellow (purple). In green is plotted the detection noise for iMIGA.}
\label{fig:allanstd}
\end{figure}

From Eq.~(\ref{eq:allanVar}) we then plot the Allan variance of the averaged atomic phase from GGN as a function of the integration time $m \, T_c$ (see Fig.~\ref{fig:allanstd}). Such calculations are done for the iMIGA configuration considering the contributions of atmospheric and seismic GGNs and the detection limit set by atom shot noise. Remarkably, we observe that fluctuations of the atomic phase induced by atmospheric GGN in noisy weather condition (purple curve) will be observable by averaging measurements over only a few seconds. Resolving fluctuations due to seismic activity would require averaging for longer periods, of the order of a few tens of seconds; long term stability commonly achieved by cold atom experiments makes this entirely within reach.
Looking at the low frequency approximation ($\omega\ll 1/m T_c$) of the transfer function (Eq.~(\ref{eq:transferGGN})) that is
\begin{equation}\label{eq:lowfreqtrans}
H_m=2k_L^2n^2m^2T_c^2T^4\omega^2 \, ,
\end{equation}
we can get an intuitive understanding of why the Allan variance of atom phase from GGN grows rapidly as a function of the integration time:
at low frequency, the weight of the low frequency components of $S_{\Delta a_x}$ in the integral is proportional to $m^2$ and $\sigma_{\psi}$ is dominated by those contributions as $S_{\Delta a_x}$ grows faster than $\omega^{-2}$.

\begin{figure}[htb]
\includegraphics[width=8.6cm]{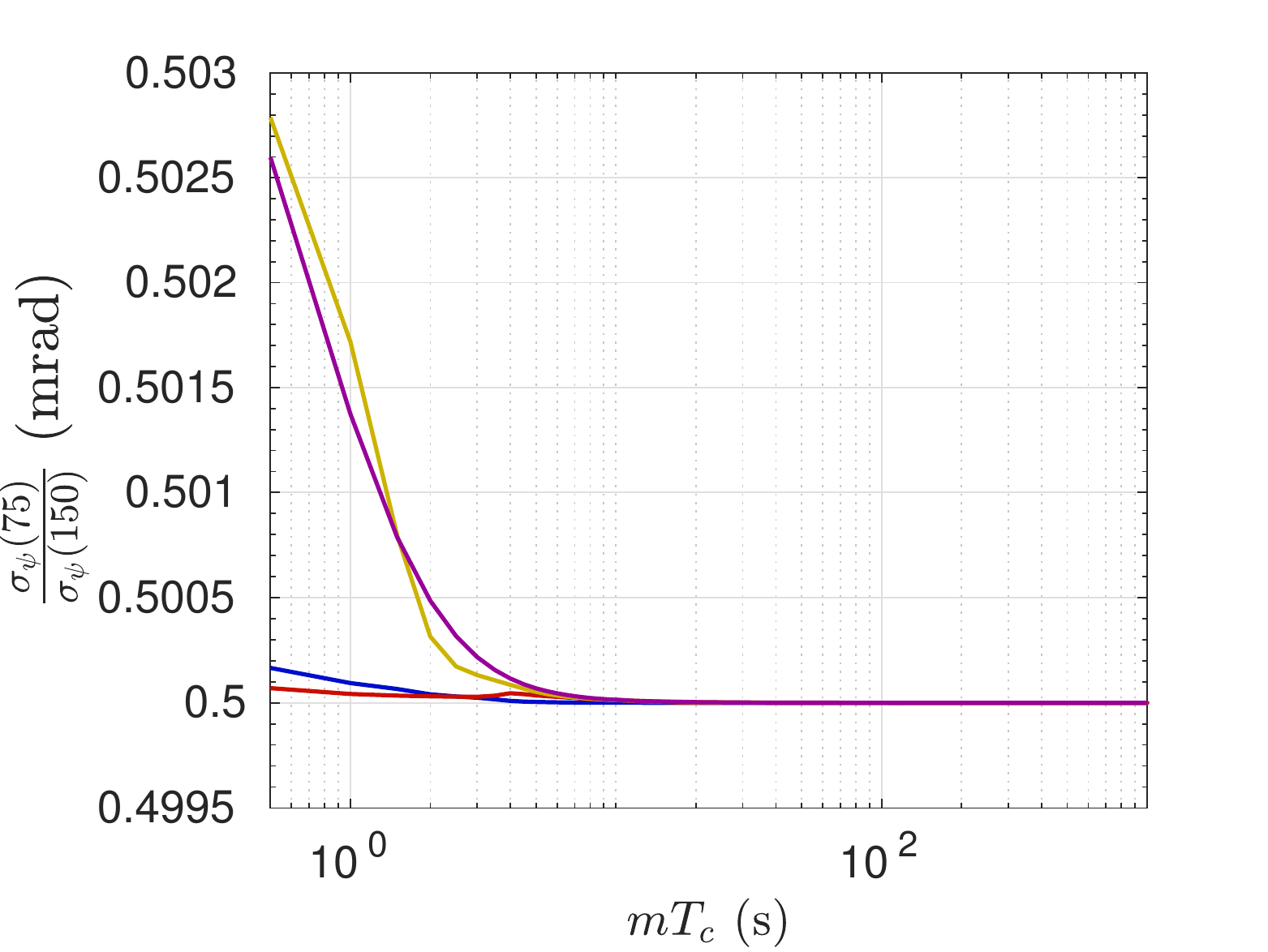}
\caption{Ratio of the Allan standard deviations of the averaged atomic phase for data set that would be simultaneously acquired from the two gradiometers available in MIGA (baselines 150 and 75 m).}
\label{fig:allanstdratio}
\end{figure}

Integrating data would give access to GGN signals from atom gradiometry measurements. The Allan variance of the averaged atomic phase from GGN could be compared with the projected values obtained from simultaneous measurements of atmospheric pressure or local acceleration, which would enable to test the models presented in Sec.~\ref{sec:seismic} and ~\ref{sec:infraS}. Remarkably, comparing the Allan variance of the different gradiometric signals of MIGA could also validate some properties of the models without the need of extra seismic or atmospheric measurements.
As the integration parameter~$m$ increases, the low frequency components of $S_{\Delta a_x}$ become dominant in the sum of Eq.~(\ref{eq:allanVar}) and the approximations of Eqs.~(\ref{eq:approxChiR},~\ref{eq:approxChiI}) become valid. Which means that averaged atomic phase $\sigma_{\psi}$ of each gradiometer will tend to be proportional to its length. This effect was numerically tested (see Fig.~\ref{fig:allanstdratio}) by calculating the ratio of the Allan standard deviations of the averaged atomic phase from GGN for the two gradiometers available in MIGA (baselines 150 and 75 m). After 10 cycles, the ratio converges to the ratio of baselines by less than one per thousand.
\section*{Conclusion}

In this paper we have modelled the effect of the local gravity field fluctuations on the MIGA underground AI and shown that the antenna will allow a study of these fluctuations. We projected the limitation of strain sensitivity from seismic and atmospheric GGN using environmental data recorded at the LSBB low noise laboratory, the location of MIGA. We have shown that the dominant contribution linked to ambient pressure variations outside the LSBB facility could become directly measurable for characteristic frequencies under 100~mHz.We also demonstrated that the integration of the antenna data on timescales of the order of a few tens of second could enable access to GGN signals with strong SNR: data integration made possible thanks to the intrinsic long term stability of cold atom experiments. This last method could refine current GGN models by comparing the measured Allan variance of the averaged atomic phase with its expected value from projection of external seismic and pressure measurements; or by comparing the Allan variance of the different MIGA gradiometers. MIGA could provide relevant GGN characterization for state-of-the art GW detectors, both in construction (KAGRA~\cite{Aso2013}) and planned (Einstein Telescope~\cite{Punturo2010}), in relation to optimal site choice and measurement protocol for background GGN. It will constitute a study case for future sub-Hz GW detectors based on AI networks for the reduction of GGN~\cite{Chaibi2016}.

\section*{Acknowledgments}
The authors would like to thank S.~Pelisson and J.~Harms for their contribution to the early stage of this study. This work was realized with the financial support of the French State through the ``Agence Nationale de la Recherche'' (ANR) in the frame of the ``Investissement d'Avenir'' programs: Equipex MIGA (ANR-11-EQPX-0028), and IdEx Bordeaux - LAPHIA (ANR-10-IDEX-03-02). This work was also supported by the r{\'e}gion d'Aquitaine (project IASIG-3D). We also aknowledge support from the CPER LSBB2020 project; funded by the ``r{\'e}gion PACA'', the ``d{\'e}partement du Vaucluse'', the MIGA Equipex and the ``FEDER PA0000321 programmation 2014-2020''. We also thank the ``P{\^o}le de comp{\'e}titivit{\'e} Route des lasers-- Bordeaux'' cluster for support. G.L. thanks DGA for financial support.

\bibliographystyle{iopart-num}
\bibliography{main}
\end{document}